\documentstyle[12pt,epsf]{article}
\newlength{\pubnumber} 

\catcode`\@=11
\@addtoreset{equation}{section}

\def\section{\@startsection{section}{1}{\z@}{3.5ex plus 1ex minus .2ex}
 {2.3ex plus .2ex}{\large\bf}}
\def\subsection{\@startsection{subsection}{2}{\z@}{2.3ex plus .2ex}
 {2.3ex plus .2ex}{\bf}}

\input epsf

\begin{document}

\begin{titlepage}
\samepage{
\setcounter{page}{1}
\rightline{LPTENS--06/22}
\rightline{LTH--700}

\rightline{\tt hep-th/0606144}
\rightline{June 2006}
\vfill
\begin{center}
 {\Large \bf 
Chiral family classification of \\
fermionic $Z_2\times Z_2$ heterotic orbifold models }
\vfill {\large Alon E. Faraggi$^{1}$,
Costas Kounnas$^{2}$\footnote{Unit\'e Mixte de Recherche
(UMR 8549) du CNRS et de l'ENS
 associ\'e©e a l'universit\'e© Pierre et Marie Curie}
 and
John Rizos$^{3}$}\\
\vspace{1cm}
{\it $^{1}$ Dept.\ of Mathematical Sciences,
             University of Liverpool,
         Liverpool L69 7ZL, UK\\}
\vspace{.05in}
{\it $^{2}$ Lab.\ Physique Th\'eorique,
Ecole Normale Sup\'erieure, F--75231 Paris 05, France\\}
\vspace{.05in}
{\it $^{3}$ Department of Physics,
              University of Ioannina, GR45110 Ioannina, Greece\\}
\vspace{.025in}
\end{center}
\vfill
\begin{abstract}

\noindent
Free fermionic construction of four dimensional string vacua,
are related to the $Z_2\times Z_2$ orbifolds at special points in the moduli
space, and yielded the most realistic three family string models
to date. Using free fermionic construction
techniques we are able to classify more than
$10^{10}$ string vacua by the net family and anti--family number.
Using  a montecarlo technique we find a
bell shaped distribution that peaks at vanishing net
number of chiral families. We also observe that $\sim15\%$ of the models
have three net chiral families.
In addition to mirror symmetry we find that the distribution exhibits
a symmetry under the exchange of (spinor plus anti--spinor)
representations with vectorial representations.

\end{abstract}
\smallskip}
\end{titlepage}

\setcounter{footnote}{0}

\def\beq{\begin{equation}}
\def\eeq{\end{equation}}
\def\beqn{\begin{eqnarray}}
\def\eeqn{\end{eqnarray}}

\def\no{\noindent }
\def\nolabel{\nonumber }
\def\ie{{\it i.e.}}
\def\eg{{\it e.g.}}
\def\half{{\textstyle{1\over 2}}}
\def\third{{\textstyle {1\over3}}}
\def\quarter{{\textstyle {1\over4}}}
\def\sixth{{\textstyle {1\over6}}}
\def\m{{\tt -}}
\def\p{{\tt +}}

\def\Tr{{\rm Tr}\, }
\def\tr{{\rm tr}\, }

\def\slash#1{#1\hskip-6pt/\hskip6pt}
\def\slk{\slash{k}}
\def\GeV{\,{\rm GeV}}
\def\TeV{\,{\rm TeV}}
\def\y{\,{\rm y}}
\def\SM{Standard--Model }
\def\SUSY{supersymmetry }
\def\SSSM{supersymmetric standard model}
\def\vev#1{\left\langle #1\right\rangle}
\def\l{\langle}
\def\r{\rangle}
\def\o#1{\frac{1}{#1}}

\def\Htw{{\tilde H}}
\def\chibar{{\overline{\chi}}}
\def\qbar{{\overline{q}}}
\def\ibar{{\overline{\imath}}}
\def\jbar{{\overline{\jmath}}}
\def\Hbar{{\overline{H}}}
\def\Qbar{{\overline{Q}}}
\def\abar{{\overline{a}}}
\def\alphabar{{\overline{\alpha}}}
\def\betabar{{\overline{\beta}}}
\def\tautwo{{ \tau_2 }}
\def\thetatwo{{ \vartheta_2 }}
\def\thetathree{{ \vartheta_3 }}
\def\thetafour{{ \vartheta_4 }}
\def\ttwo{{\vartheta_2}}
\def\tthree{{\vartheta_3}}
\def\tfour{{\vartheta_4}}
\def\ti{{\vartheta_i}}
\def\tj{{\vartheta_j}}
\def\tk{{\vartheta_k}}
\def\calF{{\cal F}}
\def\smallmatrix#1#2#3#4{{ {{#1}~{#2}\choose{#3}~{#4}} }}
\def\ab{{\alpha\beta}}
\def\Minv{{ (M^{-1}_\ab)_{ij} }}
\def\bone{{\bf 1}}
\def\ii{{(i)}}
\def\V{{\bf V}}
\def\N{{\bf N}}

\def\b{{\bf b}}
\def\S{{\bf S}}
\def\X{{\bf X}}
\def\I{{\bf I}}
\def\mb{{\mathbf b}}
\def\mS{{\mathbf S}}
\def\mX{{\mathbf X}}
\def\mI{{\mathbf I}}
\def\balpha{{\mathbf \alpha}}
\def\bbeta{{\mathbf \beta}}
\def\bgamma{{\mathbf \gamma}}
\def\bxi{{\mathbf \xi}}

\def\t#1#2{{ \Theta\left\lbrack \matrix{ {#1}\cr {#2}\cr }\right\rbrack }}
\def\C#1#2{{ C\left\lbrack \matrix{ {#1}\cr {#2}\cr }\right\rbrack }}
\def\tp#1#2{{ \Theta'\left\lbrack \matrix{ {#1}\cr {#2}\cr }\right\rbrack }}
\def\tpp#1#2{{ \Theta''\left\lbrack \matrix{ {#1}\cr {#2}\cr }\right\rbrack }}
\def\l{\langle}
\def\r{\rangle}
\newcommand{\cc}[2]{c{#1\atopwithdelims[]#2}}
\newcommand{\nn}{\nonumber}


\def\inbar{\,\vrule height1.5ex width.4pt depth0pt}

\def\IC{\relax\hbox{$\inbar\kern-.3em{\rm C}$}}
\def\IQ{\relax\hbox{$\inbar\kern-.3em{\rm Q}$}}
\def\IR{\relax{\rm I\kern-.18em R}}
 \font\cmss=cmss10 \font\cmsss=cmss10 at 7pt
\def\IZ{\relax\ifmmode\mathchoice
 {\hbox{\cmss Z\kern-.4em Z}}{\hbox{\cmss Z\kern-.4em Z}}
 {\lower.9pt\hbox{\cmsss Z\kern-.4em Z}}
 {\lower1.2pt\hbox{\cmsss Z\kern-.4em Z}}\else{\cmss Z\kern-.4em Z}\fi}

\def\AEF{A.E. Faraggi}
\def\NPB#1#2#3{{\it Nucl.\ Phys.}\/ {\bf B#1} (#2) #3}
\def\PLB#1#2#3{{\it Phys.\ Lett.}\/ {\bf B#1} (#2) #3}
\def\PRD#1#2#3{{\it Phys.\ Rev.}\/ {\bf D#1} (#2) #3}
\def\PRL#1#2#3{{\it Phys.\ Rev.\ Lett.}\/ {\bf #1} (#2) #3}
\def\PRT#1#2#3{{\it Phys.\ Rep.}\/ {\bf#1} (#2) #3}
\def\MODA#1#2#3{{\it Mod.\ Phys.\ Lett.}\/ {\bf A#1} (#2) #3}
\def\IJMP#1#2#3{{\it Int.\ J.\ Mod.\ Phys.}\/ {\bf A#1} (#2) #3}
\def\nuvc#1#2#3{{\it Nuovo Cimento}\/ {\bf #1A} (#2) #3}
\def\RPP#1#2#3{{\it Rept.\ Prog.\ Phys.}\/ {\bf #1} (#2) #3}
\def\etal{{\it et al\/}}

\hyphenation{su-per-sym-met-ric non-su-per-sym-met-ric}
\hyphenation{space-time-super-sym-met-ric}
\hyphenation{mod-u-lar mod-u-lar--in-var-i-ant}


\setcounter{footnote}{0}
\section{Introduction}

The four dimensional superstring vacua based on
the free fermionic construction \cite{fff}
are $Z_2\times Z_2$ toroidal orbifolds at
special points in the moduli space \cite{z2z21, z2z22}.
The correspondence of the free fermionic
point in the moduli space to T--self--dual points with maximally enhanced
symmetries suggests that symmetry enhancement and self--duality play a role
in the string vacuum selection.
Furthermore, the three generation heterotic string models \cite{ffm}
in the free fermionic formulation \cite{fff}
are the most realistic string models constructed to date.
The phenomenological appeal,
and the theoretical considerations, motivate the elaborate study
of this class of string compactifications.

We have therefore embarked in ref. \cite{fknr}
on a complete classification of symmetric $Z_2\times Z_2$ free fermionic
orbifold models, according to the chiral content and the four dimensional
matter gauge group. Thanks to the observation \cite{fknr}
that the twisted matter in the
models does not depend on the moduli, their chirality classification
can be carried out at the free fermionic
point of the moduli space.
Thus, this enables utilizing free fermionic techniques, which
allows an algorithm adaptable to a computer program.
Resorting to
the well known relations in two dimensions between fermionic and bosonic
currents, one can find the translation of the partition function in the
bosonic and fermionic representation, and this for any
arbitrary point of the moduli space.
Hence, the free fermionic analysis enables the chirality classification
of all symmetric, as well as the asymmetric, $Z_2\times Z_2$ orbifolds.
Thus, the free fermionic formalism provides powerful tools for the complete
classification of $Z_2\times Z_2$ perturbative string orbifolds.

The general techniques for carrying out such a classification in the
free fermionic language were developed in ref. \cite{gkr}
for type II string, and applied in ref. \cite{fknr}
for the classification of heterotic chiral $Z_2\times Z_2$
models. The analysis in ref. \cite{fknr} was performed with
respect to a subclass of the models. The $Z_2\times Z_2$ orbifold
of a six dimensional compact manifold contains three twisted
sectors. In the heterotic string each one of these sectors
may, or may not, a priori
(prior to application of the Generalized GSO (GGSO) projections),
give rise to spinorial representations.
Models that may produce, a priori, spinorial representations
from all three twisted sectors were dubbed $S^3$ models. This
class was classified in ref. \cite{fknr}.

It is also possible that the spinorial representations are not present in a
given twisted plane.
Thus, generically we may classify the models in four distinct classes:
$S^3$, $S^2V$, $SV^2$ and $V^3$ classes of models with spinorial
representations arising from three, two, one or none of the twisted sectors,
respectively. The aim of this work is to go beyond the analysis of
ref. \cite{fknr} and complete the chirality classification
of the $Z_2\times Z_2$ symmetric orbifolds.

In this process we find several surprising results. One can
envision performing the classification by removing or modifying some
vectors from the basis set which was utilized in ref. \cite{fknr},
in such a way that only two, one or none of the twisted sectors
produces massless spinorial representations. This method was pursued
in ref. \cite{nooij}. However, it is found that
the entire sets of $S^3,~S^2V,~SV^2$ and $V^3$ classes of models
are produced by
working with the basis set of ref. \cite{fknr} for certain choices
of the 
one--loop GGSO projection coefficients (discrete torsions). 
This result arises from
theta function identities in the one--loop partition function,
which we exhibit in the simplest case as an illustration.
Hence, these identities allow a complete classification of
the free fermionic $Z_2\times Z_2$ orbifold models using
a single set of boundary condition basis vectors and varying the
phases. This enables a systematic analysis of the models and
the production of algebraic formulas for the main features of
the models like the number of spinorial, anti--spinorial and vectorial
representations.
While in the past the studies of phenomenologically relevant
free fermionic
string models has been confined to isolated examples, the new
methodology allows us to scan a range of over $10^{16}$ models,
and therefore obtain vital insight into the properties
of the entire space of $Z_2\times Z_2$ orbifold vacua.
In this paper we present the main outline of the analysis
and highlights of the results.
Further details and results
will be reported in a forthcoming publication.

\section{Review of the classification method}

In the free fermionic formulation the 4-dimensional heterotic string,
in the light-cone gauge, is described
by $20$ left--moving  and $44$ right--moving two dimensional real
fermions \cite{fff}.
A large number of models can be constructed by choosing
different phases picked up by   fermions ($f_A, A=1,\dots,44$) when transported
along
the torus non-contractible loops.
Each model corresponds to a particular choice of fermion phases consistent with
modular invariance
that can be generated by a set of  basis vectors $v_i,i=1,\dots,n$,
$$v_i=\left\{\alpha_i(f_1),\alpha_i(f_{2}),\alpha_i(f_{3}))\dots\right\}$$
describing the transformation  properties of each fermion
\begin{equation}
f_A\to -e^{i\pi\alpha_i(f_A)}\ f_A, \ , A=1,\dots,44~.
\end{equation}
The basis vectors span a space $\Xi$ which consists of $2^N$ sectors that give
rise to the string spectrum. Each sector is given by
\begin{equation}
\xi = \sum N_i v_i,\ \  N_i =0,1
\end{equation}
The spectrum is truncated by a GGSO projection whose action on a
string state  $|S>$ is
\begin{equation}\label{eq:gso}
e^{i\pi v_i\cdot F_S} |S> = \delta_{S}\ \cc{S}{v_i} |S>,
\end{equation}
where $F_S$ is the fermion number operator and $\delta_{S}=\pm1$ is the
spacetime spin statistics index.
Different sets of projection coefficients $\cc{S}{v_i}=\pm1$ consistent with
modular invariance give
rise to different models. Summarizing: a model can be defined uniquely by a set
of basis vectors $v_i,i=1,\dots,n$
and a set of $2^{N(N-1)/2}$ independent projections coefficients
$\cc{v_i}{v_j}, i>j$.

The two dimensional 
free fermions in the light-cone gauge (in the usual notation) are:
$\psi^\mu, \chi^i,y^i, \omega^i, i=1,\dots,6$ (real left-moving fermions)
and
$\bar{y}^i,\bar{\omega}^i, i=1,\dots,6$ (real right-moving fermions), 
$\psi^A, A=1,\dots,5$, $\bar{\eta}^B, B=1,2,3$, $\bar{\phi}^\alpha,
\alpha=1,\ldots,8$ (complex right-moving fermions).
The class of models under investigation,
is generated by a set $V$ of 12 basis vectors
$$
V=\{v_1,v_2,\dots,v_{12}\},
$$
where
\begin{eqnarray}
v_1=1&=&\{\psi^\mu,\
\chi^{1,\dots,6},y^{1,\dots,6}, \omega^{1,\dots,6}| \nn\\
& & ~~~\bar{y}^{1,\dots,6},\bar{\omega}^{1,\dots,6},
\bar{\eta}^{1,2,3},
\bar{\psi}^{1,\dots,5},\bar{\phi}^{1,\dots,8}\},\nn\\
v_2=S&=&\{\psi^\mu,\chi^{1,\dots,6}\},\nn\\
v_{2+i}=e_i&=&\{y^{i},\omega^{i}|\bar{y}^i,\bar{\omega}^i\}, \
i=1,\dots,6,\nn\\
v_{9}=b_1&=&\{\chi^{34},\chi^{56},y^{34},y^{56}|\bar{y}^{34},
\bar{y}^{56},\bar{\eta}^1,\bar{\psi}^{1,\dots,5}\},\label{basis}\\
v_{10}=b_2&=&\{\chi^{12},\chi^{56},y^{12},y^{56}|\bar{y}^{12},
\bar{y}^{56},\bar{\eta}^2,\bar{\psi}^{1,\dots,5}\},\nn\\
v_{11}=z_1&=&\{\bar{\phi}^{1,\dots,4}\},\nn\\
v_{12}=z_2&=&\{\bar{\phi}^{5,\dots,8}\}.\nn
\end{eqnarray}
The vectors $1,S$ generate an
$N=4$ supersymmetric model. The vectors $e_i,i=1,\dots,6$ give rise
to all possible symmetric shifts of the six internal fermionized coordinates
($\partial X^i=y^i\omega^i, {\bar\partial} X^i= \bar{y}^i\bar{\omega}^i$),
while $b_1$ and $b_2$
defines the $Z_2\times Z_2$ orbifold twists. The remaining fermions not
affected by the action
of the previous vectors $\{S,e^i, b_i\}$ 
are $\bar{\phi}^i,i=1,\dots,8$ which normally give rise
to the hidden sector gauge group.
The vectors $z_1,z_2$ divide these eight complex fermions into two sets of
four. We stress here that the choice of $V$ is the most general set of
basis vectors, with symmetric shifts for the internal
fermions, compatible with a
Kac--Moody level one $SO(10)$ embedding.
Without loss of generality we can fix the associated projection coefficients
$$
\cc{1}{1}=\cc{1}{S}=\cc{S}{S}=\cc{S}{e_i}=\cc{S}{b_A}=-
\cc{b_2}{S}=\cc{S}{z_n}=-1,~
$$
leaving 55 independent coefficients,
\begin{eqnarray}
&&\cc{e_i}{e_j}, i\ge j, \ \ \cc{b_1}{b_2}, \ \ \cc{z_1}{z_2},\nn\\
&&\cc{e_i}{z_n}, \cc{e_i}{b_A},\cc{b_A}{z_n},
\ \ i,j=1,\dots6\,\ ,\  A,B,m,n=1,2\nn,
\end{eqnarray}
since the remaining projection coefficients are determined by modular
invariance \cite{fff}.
Each of the 55 independent coefficients can take two discrete
values $\pm1$ and thus a simple counting gives $2^{55}$
(that is approximately $10^{16.6}$) distinct models in the
class of superstring vacua under consideration.

The vector bosons from the untwisted sector generate an
$
SO(10)\times{U(1)}^3\times{SO(8)}^2
$
gauge symmetry.
Depending on the  choices of the projection coefficients,
extra gauge bosons may arise from
$$
x=1+S+\sum_{i=1}^{6}e_i+z_1+z_2=\{{\bar{\eta}^{123},\bar{\psi}^{12345}}\}~
$$
changing the gauge group $SO(10)\times{U(1)}\to E_6$.
Additional gauge bosons can arise as well from the sectors
$z_1,z_2$ and $z_1+z_2$ and enhance the hidden gauge group
${SO(8)}^2\to SO(16)$ or even ${SO(8)}^2\to E_8$.
Indeed, as was shown in ref. \cite{fknr},
for particular choices of the projection coefficients
a variety of gauge groups is obtained.

The matter spectrum from the untwisted sector is common to all models
and consists of six vectors of $SO(10)$ and 12 non-Abelian gauge group
singlets. The chiral spinorial representations arise from the following 48 
twisted sectors
\begin{eqnarray}
B_{\ell_3^1\ell_4^1\ell_5^1\ell_6^1}^1&=&S+b_1+\ell_3^1 e_3+\ell_4^1 e_4 +
\ell_5^1 e_5 + \ell_6^1 e_6 \nn\\
B_{\ell_1^2\ell_2^2\ell_5^2\ell_6^2}^2&=&S+b_2+\ell_1^2 e_1+\ell_2^2 e_2 +
\ell_5^2 e_5 + \ell_6^2 e_6 \label{ss}\\
B_{\ell_1^3\ell_2^3\ell_3^3\ell_4^3}^3&=&
S+b_3+ \ell_1^3 e_1+\ell_2^3 e_2 +\ell_3^3 e_3+ \ell_4^3 e_4\nn
\end{eqnarray}
where $\ell_i^j=0,1$ and $b_3=1+S+b_1+b_2+\sum_{i=1}^6 e_i+\sum_{n=1}^2 z_n$.
These states are  spinorials of $SO(10)$ and one can obtain at maximum one
spinorial ($\bf 16$ or
$\bf {\overline{{16}}}$) per sector and thus totally 48 spinorials.
Extra non chiral matter, i.e. vectors of $SO(10)$ as well as singlets,
arise from the 
$B_{\ell_3^i\ell_4^i\ell_5^i\ell_6^i}^i+x$\ ,  $(i=1,2,3)$
twisted sectors.

This construction therefore separates the fixed points of the $Z_2\times
Z_2$ orbifold into different sectors. This enables the analysis of
the  GGSO projection on the spectrum from each individual fixed point
separately. Hence, depending on the choice of the GGSO projection
coefficients we can distinguish several possibilities for the
spectrum from each individual fixed point. For example,
in the case of enhancement
of the $SO(10)$ symmetry to $E_6$ each individual fixed point
gives rise to spinorial as well as vectorial representation of
$SO(10)$ which are embedded in the $27$ representation of $E_6$.
When $E_6$ is broken each fixed point typically will give rise
to either spinorial or vectorial representation of $E_6$. However,
there exist also rare situations, depending on the choice of
GGSO phases,
where a fixed point can yield a spinorial as well as vectorial
representation of $SO(10)$ without enhancement. The crucial point,
however, is that the GGSO projections can be written as simple
algebraic conditions, and hence the classification is amenable
to a computerized analysis.

In ref. \cite{fknr} we restricted the analysis to the case
$\cc{z_1}{z_2}=-1$. Prior to GGSO projections spinorial
representations in this construction can arise from all three
twisted sectors and we therefore referred to this class as $S^3$
models. This is somewhat of a misnomer as we discuss below.
To produce models with spinorial representations arising only from
one or two of the twisted one can contemplate modifying the
basis vectors. For example, removing $z_2$ from the set
will entail that the third twisted place produces only massive
states. Hence, this would correspond to the models dubbed as $S^2V$.
Similarly, modifying the $Z_2\times Z_2$ basis vectors $b_1$ and $b_2$
in such a way that they produce vectorial rather than spinorial
representation and removing both $z_1$ and $z_2$ from the base
would entail that only vectorial representations are generated.
An analysis along this lines was followed in ref. \cite{nooij}.
However, as a result of Jacobi theta function identities
it turns out that the analysis can be carried entirely with the
basis (\ref{basis}) and just modifying the phases. To illustrate this
correspondence
we consider the simplest possibility given by the set
$\{{1},S,x\}$.
The partition function is:
\begin{equation}
\{\theta_3^4-\theta_2^4-\theta_4^4\} \{\theta_3^6{\bar\theta}_3^{14}+
                    \theta_2^6{\bar\theta}_2^{14}+
                    \theta_4^6{\bar\theta}_4^{14}\}
                  \{{\bar\theta}_2^{8}+
                    {\bar\theta}_3^{8}+
                    {\bar\theta}_4^{8}\}.~
\end{equation}
The gauge group is $SO(28)\times E_8$.
Now consider the set $\{{1},S,x,z_1\}$.
There are now two consistent choices of the coefficient
$\cc{x}{z_1}=\pm1$. The choice $\cc{x}{z_1}=+1$ produces the
$SO(28)\times E_8$ gauge group, while the choice $\cc{x}{z_1}=-1$
produces an $SO(20)\times SO(24)$ gauge group.
Indeed, the one--loop partition function as function of
$\cc{x}{z_1}$ becomes 
\begin{eqnarray}
{1\over 2}~~\left\{\theta_3^4-\theta_2^4-\theta_4^4\right\}
   ~ \left\{\theta_3^6{\bar\theta}_3^{10}\right\}~
     \left\{
                    {\bar\theta}_3^4{\bar\theta}_2^8+
                {\bar\theta}_3^4{\bar\theta}_3^8+
                {\bar\theta}_3^4{\bar\theta}_4^8+
                            \right.
&  &           \nonumber\\
\cc{z_1}{x}         {\bar\theta}_4^4{\bar\theta}_2^8+
                {\bar\theta}_4^4{\bar\theta}_3^8+
                {\bar\theta}_4^4{\bar\theta}_4^8+
&  &  \nonumber\\
                    {\bar\theta}_2^4{\bar\theta}_2^8+
                {\bar\theta}_2^4{\bar\theta}_3^8+
\cc{z_1}{x}         {\bar\theta}_2^4{\bar\theta}_4^8    \left.
                               \right\}+&  & \cdots
 \nonumber
\end{eqnarray}
Plus two additional groups of terms with the permutation of
${\bar\theta}_3$, ${\bar\theta}_2$ and ${\bar\theta}_4$.
Fixing $\cc{z_1}{x}=+1$ and using the Jacobi identity
${\bar\theta}_3^4-{\bar\theta}_2^4-{\bar\theta}_4^4\equiv0$
reproduces the partition function of the set $\{{1},S,x\}$.
This simple example illustrates how we may obtain vacua from an
enlarged basis set $\{1,S, x, z_1\}$ identical to those obtained
from a reduced basis set $\{1,S, x\}$ for an appropriate choice
of the GGSO coefficient  $\cc{z_1}{x}$, and is the primary feature which is
exploited in our classification.

\section{Counting the twisted matter spectrum}

The counting of spinorials can proceed as follows.
For each $SO(10)$ spinorial  $B^i_{pqrs}$ in (\ref{ss})
we write down the associated projector
$P^i_{pqrs}=0,1$.  The detailed expressions for the 48 projectors are
\begin{eqnarray}
P_{\ell_3\ell_4\ell_5\ell_6}^{(1)}&=&
\frac{1}{4}\,\left(1-\cc{e_1}{B_{\ell_3\ell_4\ell_5\ell_6}^{(1)}}\right)\,
\left(1-\cc{e_2}{B_{\ell_3\ell_4\ell_5\ell_6}^{(1)}}\right)\,\nn\\
&&\frac{1}{4}\left(1-\cc{z_1}{B_{\ell_3\ell_4\ell_5\ell_6}^{(1)}}\right)\,
\left(1-\cc{z_2}{B_{\ell_3\ell_4\ell_5\ell_6}^{(1)}}\right)\nn\\
P_{\ell_1\ell_2\ell_5\ell_6}^{(2)}&=&
\frac{1}{4}\,\left(1-\cc{e_3}{B_{\ell_1\ell_2\ell_5\ell_6}^{(2)}}\right)\,
\left(1-\cc{e_4}{B_{\ell_1\ell_2\ell_5\ell_6}^{(2)}}\right)\,\nn\\
&&\frac{1}{4}\,\left(1-\cc{z_1}{B_{\ell_1\ell_2\ell_5\ell_6}^{(2)}}\right)\,
\left(1-\cc{z_2}{B_{\ell_1\ell_2\ell_5\ell_6}^{(2)}}\right)\label{proj}
\\
P_{\ell_1\ell_2\ell_3\ell_4}^{(3)}&=&
\frac{1}{4}\,\left(1-\cc{e_5}{B_{\ell_1\ell_2\ell_3\ell_4}^{(3)}}\right)\,
\left(1-\cc{e_6}{B_{\ell_1\ell_2\ell_3\ell_4}^{(3)}}\right)\,\nn\\
&&\frac{1}{4}\,\left(1-\cc{z_1}{B_{\ell_1\ell_2\ell_3\ell_4}^{(3)}}\right)\,
\left(1-\cc{z_2}{B_{\ell_1\ell_2\ell_3\ell_4}^{(3)}}\right)\nn
\end{eqnarray}
For the surviving spinorial ($P^i_{pqrs}=1$)
the  chirality (${\bf 16}$ or ${\bf \overline{16}}\,$)
is determined from the associated chirality
coefficient $X^i_{pqrs}=\pm1$, where
\begin{eqnarray}
X_{\ell_1\ell_2\ell_5\ell_6}^{(1)}&=&-\cc{b_2+
(1-\ell_5) e_5+(1-\ell_6) e_6}{B_{\ell_3\ell_4\ell_5\ell_6}^{(1)}}
\nn
\\
X_{\ell_1\ell_2\ell_5\ell_6}^{(2)}&=&-\cc{b_1+
(1-\ell_5) e_5+(1-\ell_6) e_6}{B_{\ell_1\ell_2\ell_5\ell_6}^{(2)}}
\\
X_{\ell_1\ell_2\ell_3\ell_4}^{(3)}&=&-\cc{b_1+(1-\ell_3) e_3+
(1-\ell_4) e_4}{B_{\ell_1\ell_2\ell_3\ell_4}^{(3)}}
\nn
\end{eqnarray}
Using the above results, we can easily calculate the number of 
spinorials/antispinorial per sector
\begin{eqnarray}
S_{\pm}^{(i)}&=&\sum_{pqrs} 
\frac{1\pm X^{(i)}_{pqrs}}{2} P_{pqrs}^{(i)}\ , \ i=1,2,3\label{cc}
\end{eqnarray}

The counting of $SO(10)$ vectorials can proceed in a similar way.
For each vectorial generating sector
$B^i_{pqrs}+x$ the associated projector  $\tilde{P}^i_{pqrs}$ is 
obtained from (\ref{proj})
using the replacement $B^i_{pqrs}\to B^i_{pqrs}+x$.
Since there is no chirality in this case the number of
vectorials per sector is just the sum of the projectors
\begin{eqnarray}
V^{(i)}=\sum_{pqrs} {\tilde P}_{pqrs}^{(i)}
\end{eqnarray}

The total vectorial ($V$) and net spinorial ($S$) numbers are
\begin{eqnarray}
V=\sum_{i=1}^3 V^{(i)}
\end{eqnarray}
and
\begin{eqnarray}
S=\sum_{i=1}^3 S_{+}^{(i)}-S_{-}^{(i)}
\end{eqnarray}

The mixed projection coefficients entering the above formulas can be
decomposed in terms of the
independent phases $\cc{v_i}{v_j}, i>j$.
After some algebra we come to the conclusion that for the counting of the
spinorial/antispinorial and
vectorial $SO(10)$ states the phases $\cc{e_i}{e_i}, i=1,\dots,6$,
$\cc{z_A}{z_A}, A=1,\dots,2$,
$\cc{b_I}{b_I}, I=1,\dots,2$ as well as $\cc{e_3}{b_1}, \cc{e_4}{b_1},
\cc{e_1}{b_2}, \cc{e_2}{b_2}$
are not relevant. Moreover the phase $\cc{b_1}{b_2}$ is related to the
total chirality flip.
This leaves a set of 40 independent phases which is still too
large for a manageable computer analysis. We therefore resort
to a Monte--Carlo analysis that generates random choices of phases.
The complete classification is currently underway and will be reported
in a future publication. In this sense our results are based on a statistical
polling, with a sample that correspond to some $10^{10}$ vacua.
We have checked however that increasing the size of the sample
by 10\% does not alter our results. Therefore our results, while
in some sense empirical, are expected to hold for the entire space
of vacua.

\section{The observable and hidden gauge groups}

An important step in the analysis is the determination of
the four dimensional gauge group. In particular, it is important
to determine the component of the gauge group which is identified
with the observable gauge group. Observable matter then consist
of the states that are charged with respect to this gauge group.
In our construction the sectors that produce space--time vector bosons
include:
$
G=\{0,z_1,z_2,z_1+z_2,x\}.~
$
The $0$ sector gauge bosons produce the gauge group
$
SO(10)\times U(1)^3\times SO(8)^2.~
$
Depending on the choice of the GGSO projection coefficients the
four dimensional gauge group is enhanced. Since the sectors
that produce space--time vector bosons do not break any supersymmetries
the classification of the four dimensional gauge group can be done
at the level of the $N=4$ vacuum. The addition of the supersymmetry
breaking sectors $b_1$ and $b_2$ then  breaks the $N=4$ gauge group
to the unbroken gauge group at the $N=1$ level. Some of the possibilities
at the $N=1$ level are listed in table \ref{gaugegroups}.

\begin{table}
\centering
\begin{tabular}{c}
\hline
Gauge group\\
\hline
$SO(10)\times{SO(18)}\times{U(1)^2}$\\
\hline
$SO(10)\times{SO(9)}^2\times{U(1)^3}$\\
\hline
$SO(10)^2\times{SO(9)}\times{U(1)^2}$\\
\hline
$SO(10)^3\times{U(1)}$\\
\hline
$SO(26)\times{U(1)}^3$\\
\hline
$E_6\times{U(1)^2}\times{E_8}$\\
\hline
$E_6\times{U(1)^2}\times{SO(16)}$\\
\hline
$E_6\times{U(1)^2}\times{SO(8)\times{SO(8)}}$\\
\hline
$SO(10)\times{U(1)}^3\times{E_8}$\\
\hline
$SO(10)\times{U(1)}^3\times{SO(16)}$\\
\hline
\end{tabular}
\caption{\label{gaugegroups}\it Typical enhanced gauge groups
for a generic model generated
by the basis (\ref{basis}).
}
\end{table}
Several features are noted from the list. The first five groups are 
descendants of
$SO(32)$, whereas the later five are descendants of $E_8\times E_8$. 
This gross division
is controlled by the phase $\cc{z_1}{z_2}$. An important point to note,
relevant for our classification,
is the occurrence of models with several $SO(10)$ group
factors. This arises because of the enhancement of the $0$--sector 
$SO(8)\times U(1)$
group factors to $SO(10)$.
In our analysis we define the observable gauge group to be $SO(10)$
and the chiral matter should be charged under that group.
The question arises as to which $SO(10)$ should be identified as
the observable gauge group and the subsequent classification of
the chiral and vectorial matter states with respect to that group.
In our analysis here we identity the observable $SO(10)$ symmetry as the
one which is generated by the world--sheet fermions ${\bar\psi}^{12345}$.
Investigations of other possibilities are left for future work, although
our experience with the construction of quasi--realistic string models
suggests that the spectrum with respect to other $SO(10)$ group factors
is vectorial rather than spinorial.

We further comment that in our classification the observable GUT gauge group
is always $SO(10)$. The conditions for enhancement of the gauge group
are given in ref. \cite{fknr}, eqs. (4.3-4.11). These conditions
are incorporated into our classification routine and cases in which
the $SO(10)$ symmetry is enhanced are rejected. 

We also note that our classification
is with respect to the chiral content of the models, which in the 
free fermionic models arises from the twisted sectors. The breaking
of the GUT $SO(10)$ symmetry to a subgroup, which can be further 
broken to the Standard Model gauge group, is achieved in free fermionic
models with  an additional boundary condition basis.
Free fermionic models with
quasi-realistic gauge group and chiral family content were produced
in the past \cite{ffm}. The interest in
this paper is in the global properties of a large class of compactifications
to which the quasi-realistic free fermionic models belong, but not in 
producing quasi-realistic spectrum. The classification of free fermionic
models with broken $SO(10)$ GUT symmetry will be pursued in future work.

\section{Results} 


\begin{figure}[!h]
\centerline{\epsfxsize 5.0 truein \epsfbox{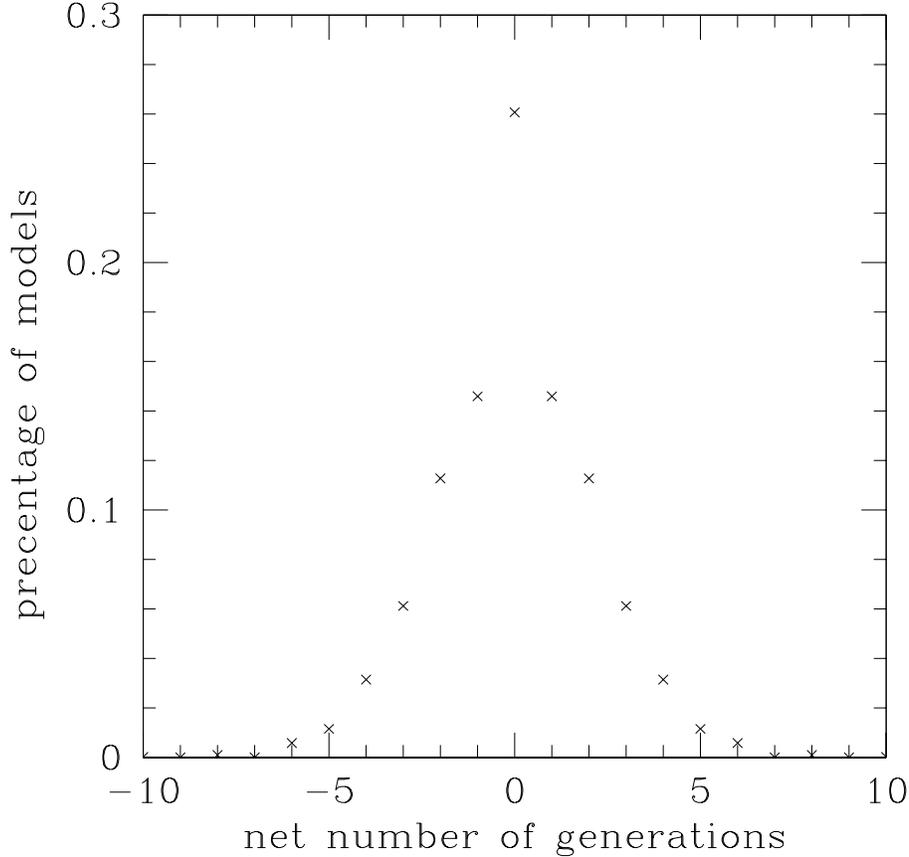}}
\label{montecarlo}
\caption[1]{
Distribution of a Monte Carlo generation of models sampling
some $10^{10}$ vacua.}
\end{figure}

\begin{figure}[!h]
\centerline{\epsfxsize 5.0 truein \epsfbox{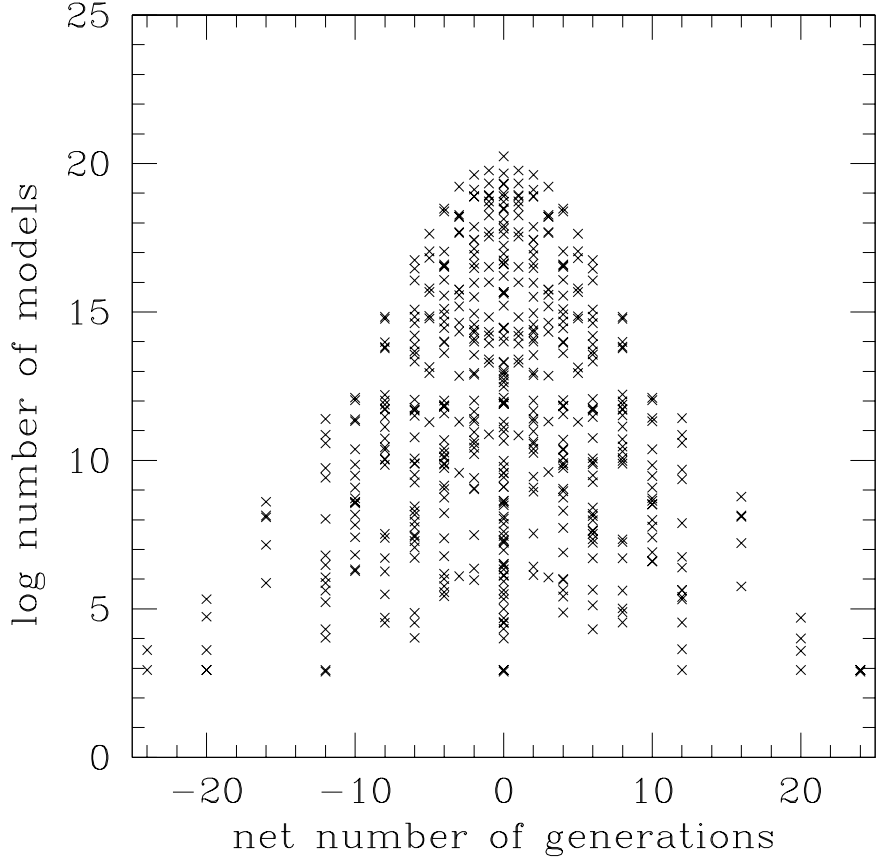}}
\caption[1]{\label{totlogscat}
Scatter plot of the logarithm of the number of models versus the
net number of chiral families.}
\end{figure}

The results of the random search are exhibited in figures
\ref{montecarlo} and \ref{totlogscat}.
The first figure shows the percentage of models with a net number of
chiral families.
The second figure exhibits the total number
of vacua on a logarithmic scale over the net number of chiral families.
In our sample the peak of the
distribution is for a vanishing net number of chiral families. About
15\% of the models contain a net number of three chiral or anti--chiral
families. By increasing the sample size by 10\% we note that these
results are not modified.

In figure \ref{totlogscat} we present scatter plot of the logarithm
of the number of models versus the net number of chiral families.
The vertical spread arises from variations in the number of vectorial
representations in the models. The peak is for vanishing net number of
chiral families, and we note that the distribution is symmetric about that
point.
The plot has a bell shape which recedes for increasing number of
chiral families. For net family number above 24 the number of occurrences
is small and are not shown on the plot. This reflects that increasing
number of chiral families requires that the configurations of the GGSO phases
attain higher symmetry and consequently the number of possibilities
decreases. Curiously, we note that models with some net number of
chiral families
do not appear on the plot. For example models with 7, 9, 11, 13, 14, 15, 17,
18, 19 net number of chiral families do not appear in our sample. Thus, these
are either very rare or forbidden altogether in symmetric $Z_2\times Z_2$
orbifolds. Additional plots and analysis of the data will be presented
in a forthcoming publication.

The symmetry about the vanishing number of chiral families is in accordance
with mirror symmetry. We note, however, an additional symmetry in the
distribution under exchange of {\it vectorial}, and {\it spinorial plus 
anti--spinorial}, representations.
The symmetry states that for a model with a given total number of spinorial
plus anti--spinorial representations there exist a corresponding model in
which
the spinorial plus anti--spinorial representations are exchanged with
vectorial representations. Ultimately, in the free fermionic language
this symmetry reflects a symmetry under a discrete exchange of some GGSO
projection coefficients. However, it may have interesting implications
in terms of the underlying geometrical data.

In conclusion, the quasi--realistic free fermionic models are among the
most realistic string models constructed to date. 
While past exploration
of these models consisted of the study of single examples, in this paper
we embarked on the investigation of the properties of the whole space
of vacua in this class. Future studies will incorporate into the analysis
further properties of the realistic models, like the assignment
of asymmetric boundary conditions. While the task is still horrendous
in terms of the shear number of vacua,
the existence of string models in this class that come close to describing
reality, and precisely where one would most expect to find them,
gives ample reason to suggest that we are on the right track.

\section{Acknowledgments}

We would like to thank Sander Nooij for collaboration at the initial
stages of this work. AEF would like to thank the Ecole Normale Sup\'eriere and
the University of Ioannina, CK would like to thank
the University of Ioannina and JR would like to thank
the Ecole Normale Sup\'eriere, for hospitality.
AEF is supported in part by PPARC under contract PP/D000416/1.
CK is supported in part by the EU under contracts 
MTRN--CT--2004--005104, MTRN--CT--2004--512194 and
ANR (CNRS--USAR) contract No 05--BLAN--0079--01 (01/12/05)
JR is supported by the program ``PYTHAGORAS''  (no. 1705 project 23)
of the Operational Program for Education and Initial Vocational
Training of the Hellenic Ministry of Education under the 
3rd Community Support Framework and the European Social Fund; 
and by the EU under contract MRTN--CT--2004--503369. 



\bigskip
\medskip

\bibliographystyle{unsrt}

\end{document}